\def\ref{\par\noindent\hang}

\def\etal{ et al.\ }

\def\km{{\rm\thinspace km}}

\def\Mpc{{\rm\thinspace Mpc }}
\def\kpc{{\rm\thinspace kpc }}

\def\s{{\rm\thinspace s}}

\def\kmps{\hbox{$\km\s^{-1}\,$}}

\def\1E{1E~1415.6+2557}

\documentstyle[11pt,aaspp4]{article}

\tightenlines%

\begin{document} 

\title{THE SLOPE OF THE CLUSTER ELLIPTICAL RED SEQUENCE: 
A PROBE OF CLUSTER EVOLUTION}

\author{Michael D. Gladders\altaffilmark{1,2}, Omar L\'opez-Cruz\altaffilmark{1,3,4},
H.K.C. Yee,\altaffilmark{1,4}}
\author{and Tadayuki Kodama\altaffilmark{5}}

\altaffiltext{1}{Department of Astronomy, University of Toronto, Toronto, ON, M5S 3H8, Canada. gladders,hyee@astro.utoronto.ca}

\altaffiltext{2}{Guest User, Canadian Astronomy Data Center, which is operated 
by the Dominion Astrophysical Observatory for the National
 Research Council of Canada's Herzberg Institute of Astrophysics. }

\altaffiltext{3}{Instituto Nacional de Astrofis{\'\i}ca, \'{O}ptica y Electr\'{o}nica(INAOE), 
Tonantzintla, Pue., M\'{e}xico. omar@inaoep.mx}

\altaffiltext{4}{Visiting Astronomer, Kitt Peak National Observatory, which is operated by AURA, Inc., under contract to the National Science Foundation.}

\altaffiltext{5}{Institute of Astronomy, Madingley Rd, Cambridge CB3 0HA,
England. kodama@ast.cam.ac.uk}

\begin{center}Accepted for publication in {\it The Astrophysical Journal}\end{center}

\vfill \begin{abstract} \noindent The current formation models for cluster elliptical galaxies which
incorporate a mechanism for the metallicity enhancement of massive ellipticals predict a change in the
observed slope of the red sequence of ellipticals as a function of redshift. This change occurs primarily 
because the metal-rich galaxies become redder faster than the metal-poorer 
galaxies with increasing age. This effect is most
pronounced within $\sim$ 4 Gyr of formation. Observations of the change of the
slope of the red sequence with redshift may thus be used to constrain the formation epoch for
galaxy clusters. We examine the red sequence of cluster ellipticals 
using publicly available HST imaging data for a set of six $0.75>$ z $>0.2$ clusters, and 
a sample of 44 Abell clusters at z $<0.15$ imaged with the KPNO 0.9 m. We compare the 
derived slopes of the red sequences with a set of cluster-elliptical evolution models
and find good agreement. We demonstrate that such a comparison provides
a useful constraint on the formation epoch for clusters, which can be made independently
 from considerations of absolute
color evolution and scatter in the red sequence. From our initial comparison
of the observed and model slopes as a function of redshift, we conclude, conservatively, 
that most of the elliptical galaxies in the
cores of clusters must form at z $>2.0$, and that these galaxies are coeval and 
passively evolving.

\end{abstract}

\vfill \it Key Words: \rm cosmology: observations - galaxies: clusters: general
 -  galaxies: formation

\eject \section{INTRODUCTION}
\noindent 

The existence of a color-magnitude relation (CMR) for field ellipticals, in which
the brighter ellipticals are in general redder,  was first noted by Baum (1959).
Locally, the elliptical galaxies in individual clusters form a red sequence, with a well--defined slope
and small scatter (Bower\etal 1992a, b).
Recent results from the Hubble Space Telescope 
demonstrate the existence of a tight red sequence, comparable in scatter 
to that observed in the red sequence of the Coma cluster, in clusters at
redshifts up to $z=0.9$ (Stanford\etal 1998, hereafter SED98; Ellis\etal 1997, hereafter E97).
Though the CMR in clusters can be interpreted at the present epoch as either an 
age effect or a metallicity effect, the existence of the red sequence at 
redshifts greater than 0.3 makes the age explanation untenable (Kodama 1997, hereafter K97).
Specifically, the existence of the red sequence at higher redshifts
indicates that cluster ellipticals are a passively evolving population in which the reddening of 
massive galaxies is the result of a mass-metallicity relation rather
than an age effect (Kodama \& Arimoto 1997; Kauffman \& Charlot 1997; K97).
\paragraph{}
 The apparent passive evolution of cluster ellipticals is broadly consistent
with models in which ellipticals form in
a monolithic collapse at high redshift, and evolve passively after this initial 
star-burst (Eggen\etal 1962). 
The origin of the mass-metallicity relation in this collapse scenario was 
first explored in detail by Arimoto \& Yoshii (1987), who included the effects 
of supernova winds (Larson 1974). The heating of the interstellar medium
by supernovae in the initial star-burst triggers the 
formation of a galactic wind when the thermal energy of the gas exceeds the gravitational
binding energy. This galactic wind ejects the gas in low-mass
galaxies more efficiently due to their shallower potential wells, resulting in a
trend of increasing metallicity with mass (e.g., Carlberg 1984a, b). The more massive galaxies
 are more likely
to retain the enriched supernova ejecta, and have star-bursts of longer duration.
\paragraph{}
The initial enrichment difference between elliptical galaxies of different masses
is manifested in the present epoch as the slope of the red sequence, which appears to be constant
from cluster to cluster (Lopez-Cruz \& Yee 1998; Lopez-Cruz 1997, hereafter LC97). However, this mass-metallicity relation
causes elliptical galaxies of different masses to
display slightly differing photometric properties with age. This differing
color evolution manifests as a change in the slope of the red sequence with redshift.
In a scenario in which elliptical galaxies in the cores of clusters form concurrently
in a monolithic collapse at high redshift, we expect the observed slope of the red sequence
to be significantly flatter at ages less than $\sim$ 4 Gyr after formation (K97; Kodama \& Arimoto 1997), when
compared to later times.

\paragraph{}
An alternative explanation for the origin of the slope of the red sequence is
provided by Kauffman \& Charlot (1997), in which elliptical galaxies are formed
hierarchically through the merger of disk systems. In this model there is again
a mass-metallicity relation for the progenitor disk systems, and the enrichment
occurs prior to the assembly of elliptical galaxies. Massive elliptical
galaxies tend to form from the more massive disk systems in hierarchical merging, 
resulting in a mass-metallicity relation for elliptical galaxies. In this
model the observed slope of the red sequence is also expected
to flatten at high redshifts, because the stellar populations in massive ellipticals
are on average younger, and become bluer relative to low-mass systems as the 
formation epoch is approached.
\paragraph{} In theory, then, the change in the slope of the red sequence with 
redshift can be used to constrain the formation epoch of the dominant stellar
population in elliptical galaxies (K97; LC97). This constraint
can be made independently of the intrinsic scatter in early-type galaxy colors about the
red sequence (Bower\etal 1992a,b; 
Stanford\etal 1995; E97; SED98), and the evolution of  
the apparent color with redshift (Arag\'{o}n-Salamanca\etal 1993; K97; E97; SED98; Kodama\etal 1997). 
The measurement of the slope offers a significant advantage
over measurements of color and scatter in that it is calibration independent.
The slope can be measured with the same precision even in the presence of large 
systematic errors in photometric zero points.
\paragraph{}
In this paper we present an analysis of the slopes of the red sequences for a total
of 50 clusters spanning the redshift range 0$<$ z $<$0.75, and demonstrate that such an 
analysis offers a powerful tool in constraining the formation epoch of elliptical galaxies
in clusters. Our data are drawn from 
two sources: a subset of the imaging survey of LC97 of 45 Abell
clusters at z $<$0.2, and six clusters at z $>$0.2 drawn from the Hubble Space Telescope (HST)
 archive. 
We consider only the interior 0.5 \Mpc of each cluster, and select early-type
galaxies in the high redshift sample using an automated galaxy classifier based
on central concentration (c.f. Abraham\etal 1994). In \S 2, we define the
imaging dataset in detail, and discuss the galaxy photometry and morphology parameters 
derived from
these data. In \S 3, we present our analysis of the red sequence slopes in these
clusters, and compare these results to the models of K97 in \S 4. In \S 5 we
summarize our conclusions, and comment on the applicability of this analysis to
even higher redshift observations. We use $H_{0}$=50, $\Omega_{{\it m}}$=1.0, and $\Omega_{\Lambda}$=0.0
throughout this paper, unless otherwise noted.

\section{THE DATA}

\subsection{The KPNO Sample}
\noindent The low-redshift data for this paper are taken from the KPNO 0.9 m survey of
45 Abell clusters of LC97. This is a three-color ($B$,$R$, and $I$) wide-field (23'$\times$23') 
survey of high galactic latitude
($|b|\geq$30), x-ray luminous clusters at z $\leq$0.2. Details of 
the data reduction, and
a more comprehensive analysis of the CMR for each cluster can be found in Lopez-Cruz \&
Yee (1998). 
In this paper, we use only the {\it B} and {\it R} photometry for all
but one of these clusters to derive the color-magnitude diagram (CMD) of the inner 0.5 \Mpc
radius of each cluster. We have excluded the highest redshift cluster (A665, $z=0.18$) in the survey,
as the data do not provide accurate photometry for
all but the brightest members of this cluster. The remaining 44 clusters have been binned into
five sub-samples in redshift; the identification of the clusters in each sub-sample is 
given in Table 1.

\subsection{The HST Sample}
\noindent The HST data used in this paper have all been drawn from the HST archive.
We selected six clusters for this study, requiring that each have deep Wide-Field
Planetary Camera 2 (WFPC2) imaging in at least two band-passes, and spanning as large
a redshift range as possible within this constraint. The specifics of
 these observations 
are presented in Table 2. In all cases, the F814 image is the deepest, and is
complemented by a deep F555 image for four clusters at z$<0.6$, and an F606 image for the two
at z$\approx$0.75. The data have been processed using the Canadian Astronomical
Data Centre standard WFPC2 pipeline, utilizing the most up-to-date calibration
frames available. We have not attempted to correct for charge transfer efficiency effects
(c.f. Holtzman\etal 1995), as this effect is position dependent and has little
effect on colors measured between two frames at the same position. The individual 
exposures for each field have been shifted and combined using a sigma-clipping algorithm 
to reject cosmic rays and hot pixels. For clusters where there are no offsets between the
frames, a bad pixel map has been used to eliminate hot pixels from consideration in
the photometry and morphological analysis.

\subsection{Photometry}
\noindent The photometry for both the KPNO and HST imaging data has been computed using the
Picture Processing Package (PPP; Yee 1991). PPP is an integrated photometry system which
performs object finding and computes photometry parameters based on an analysis of the curve of
growth for each object. In all cases, the redder image for each cluster was used to 
find objects, and each frame has been examined individually to confirm the reliability of
the object finding. PPP reliably separates stars from galaxies by reference to the curve
of growth of a set of reference stars, and we have excluded all objects identified by PPP 
as stars. This classification has been visually
checked for all bright objects in the KPNO data, as bright stars can be misclassified
due to variations in the PSF at the edges of the field.
The magnitudes used here are total magnitudes (see Yee 1991 and LC97 for details).
A standard metric aperture of 11 \kpc was used when computing colors for the KPNO data
at z$<0.06$, and all the HST data. For the KPNO data at z$>0.06$, we have used a fixed
observed aperture of 6'', corresponding to $\approx$ 21 \kpc at z=0.15. The larger observed apertures
have been used in the KPNO data for clusters at z$>0.06$, because the 11 \kpc metric aperture is too
small relative to the seeing disk to allow for a reliable color determination. However, 
the difference in apertures
is insignificant, as the color gradients in elliptical galaxies are very small 
(Peletier\etal 1990), and the variation this induces in the expected slope of the red sequence
is negligible (K97).

\paragraph{} The KPNO imaging data have the added complication of containing a large number
of clusters with massive cD galaxies. These galaxies have been treated individually by
isophote modeling (Brown 1997), and removed from the cluster images prior
to the computation of photometry on all other objects in the field. This avoids the difficulty
of computing the background for objects within the extended cD haloes. 

\subsection{Morphological Classification}
\noindent The CMDs for the higher redshift clusters in our sample are significantly contaminated by
field galaxies along the line of sight. As the slope of the red sequence is well
 determined only when this background
contamination is minimal, we have used morphological classification to select only the
early-type population. Dressler\etal (1994) have 
demonstrated that HST images have sufficient resolution to allow reliable classification of
galaxy morphologies at high redshift. More recently, E97 have used this to identify
the E/S0 population in three clusters at z$\approx$0.55. In this paper we attempt a similar
procedure to isolate the early-type galaxies in the HST sub-sample, with the aim of improving the
signal of the red sequence in the CMD. Following the prescription of Abraham\etal (1994), we
use central concentration as our primary classification tool. The concentration index, $C$, is 
defined as the ratio of the flux in the central 1/3 of the object, compared to the total flux
contained within the 2-$\sigma$ isophote. We chose to use the object magnitude, rather than
surface brightness (as adopted by Abraham\etal), as our secondary classification criterion in the
``classification plane'' (see Figure 1a).
\paragraph{} To isolate the red sequence in the HST subsample most optimally, we proceeded as follows: First,
we made a cut in the concentration-magnitude plane, and examined the CMD resulting
from concentrated objects selected by this cut. In all cases, the red sequence was apparent.
We then defined the ``red sequence region'' in the CMD, and examined the position of {\it all}
objects in the region, in the classification plane. We next made a new cut in the classification
plane which best separated the two groups (see Figure 1a). 
Objects which passed this second cut were then 
individually examined, and galaxies which were obviously not E/S0 galaxies were excluded. This
visual examination excluded $\approx$ 10\% of the remaining objects. These excluded objects were
typically spiral galaxies with concentrated central bulges, or
crowded objects. The final sub-set of concentrated objects is composed predominantly
of early-type galaxies, and in all cases, the visibility of the red sequence in the cluster
CMD was strongly enhanced (see Figure 1b).

\subsection{Color-Magnitude Diagrams}
\paragraph{}As we are interested in only the early-type cluster members, we have used
only the inner 0.5 \Mpc projected radius of each cluster to construct its CMD. 
This reduces the relative
contamination of the CMD by field objects (particularly important for the highest redshift
clusters), as we are examining the density peak of the cluster. Furthermore, the cores of
clusters are in general dominated by early-type galaxies (Dressler 1980). Recent 
analysis of several clusters at $z\approx0.55$ (Dressler\etal 1997) shows a paucity of
S0 galaxies in the cores of rich clusters, relative to a well established, centrally
concentrated population of ellipticals. We have made no attempt to segregate the
E and S0 populations morphologically, but we expect that the restricted field size used here
favors elliptical galaxies over S0 galaxies. Notably, Bower\etal (1992b) and E97 found no 
discernible difference between the slope and zero-point of the red sequences of cluster E and S0 galaxies 
at z$\approx$0 and z$\approx$0.55 respectively,
 so the inclusion of some S0 galaxies is unlikely to have a significant effect. Ellingson\etal (1998) 
have performed a principal-component analysis of the spectra of cluster 
members for 14 clusters in the CNOC1 survey, and find an increase in the 
Balmer component of the spectra only at radii
comparable to, or greater than, $r_{200}$, the radius at which the average interior density for
the cluster is $\approx$200$\rho_{c}$, where $\rho_{c}$ is the
critical density. Ellingson\etal interpret this increase in the 
Balmer component as the signature of recent star-formation truncated by infall into the 
inter-cluster medium. Following Carlberg\etal (1997), the expected value of $r_{200}$
for a cluster at z=0.75, with a velocity dispersion $\sigma_{c}=800$
\kmps, is $\approx$1 \Mpc, well beyond the 0.5 \Mpc radius used here. The observation that there is no
significant spectroscopic evidence for recent star formation in S0 galaxies in the cores of clusters 
is also consistent with Van Dokkum\etal (1998), who found a blueing of the S0 population
in CL1358+62 (z=0.33) only at radii greater than 0.7 {\it h}$^{-1}_{50}$ \Mpc. We thus expect our
sample to be dominated by elliptical galaxies, with a contribution from quiescent S0
galaxies with no remnant signature of star-formation. Moreover, the inclusion of a small population of 
quiescent S0 galaxies is not expected to significantly affect our slope measurements, because an S0 galaxy
which ceases star formation in the disk will enter the red sequence within a few Gyr (Bower, Kodama,
\& Terlevich 1998) and be indistinguishable from an elliptical.

\paragraph{}
The 44 clusters in the KPNO sub-sample have been subdivided into five redshift groups, 
and the CMDs ($B-R$ versus {\it R}, in all cases) for all clusters in each group have been combined.
Three corrections have been applied to
the individual CMDs prior to combination. First, we have applied galactic
extinction corrections, using the maps and tabulations of Burnstein \& Heiles (1982, 1984).
Second, we have corrected the colors using the fit to the red sequence $B-R$ color versus
redshift of LC97 -- this is essentially an empirically determined color K-correction.
Third, we have corrected the {\it R}-band magnitudes to a common absolute magnitude, considering only
cosmological effects with no K-correction. This correction to a common absolute magnitude
is sufficient, as the total spread in redshift in each group is only $\approx$ 0.02. 
\paragraph{}
The CMDs for the two pairs of the highest redshift clusters with HST data have also been combined. To combine
J1888.16CL with CL0016+1609, we have applied the color correction from E97 of --0.025 mag to
the CMD of J1888.16CL. We have combined the CMDs of GHO1322+3027 and GHO1322+3114 directly, as
the galactic extinction for these two clusters is the same, and their redshifts are only marginally
different.
\paragraph{}
The final combined CMDs for both subsets of the data are shown in Figures 2a-h. 
In each combined set the red sequence is apparent. There is significant contamination of 
the CMD by non-red sequence objects in the KPNO sub-sample, but the large number of clusters
in each group adequately compensates for this.

\section{RED SEQUENCE SLOPES}
\noindent
We have computed the red sequence slopes for our CMD sample using both standard linear
regression, and a method based on minimizing the absolute deviation (Press\etal 1992).
The minimization of the least absolute deviation (LAD) is expected to be more robust
in the presence of outliers. The estimated color errors were used to weight the data for
all fits.
We have investigated the two above methods in conjunction with
 $\sigma$ clipping about 
the red sequence to exclude
obvious outliers, and have found that linear regression plus an iterated 3-$\sigma$ clip
provided the most stable results. All results discussed below were produced using this
fitting technique. 
\paragraph{} Specifically, the fits to each CMD reported in Table 3 were computed as
 follows: First, a limiting magnitude was selected for each CMD, under the condition that
the red sequence be apparent up to this limit. An initial estimate of the 
fit was then made in
a visually selected sub-region of the CMD containing the red sequence. 
The CMD was rectified by this fit, and the color distribution of {\it all} objects in the 
color-corrected CMD was computed. A Gaussian was then fit to the color peak corresponding to the
red sequence, and only points less than 3$\sigma$ from the peak were used to construct a new
red sequence fit. This process was iterated until convergence on a final solution.
We have investigated the sensitivity of this solution to reasonable choices of limiting magnitude
and initial fit, and found no significant deviations from our final fits in all cases.
The final errors on the slopes reported in Table 3 were estimated by a boot-strap
analysis of our fitting procedure. For CMDs comprising data from more than one cluster, 
the effective redshift is estimated as the mean of the cluster redshifts in the subsample,
weighted by the number of galaxies from each cluster in the final
fitted region of the CMD.

\section{CLUSTER FORMATION TIME}
\noindent We have chosen to compare our derived red sequence slopes
to the models of K97. We make no detailed comparison to 
the hierarchical-clustering based model of Kauffmann \& Charlot (1997), because 
they do not provide sufficiently detailed predictions of the red sequence slopes 
in the different band-passes used here to allow such a comparison.

\paragraph{} The models of K97 are normalized to reproduce the slope of the 
red sequence in the Coma cluster (A1656) observed by Bower\etal (1992b), and we use
that normalization here.
We use the models of K97 to correct our observed
slopes in various colors to a common color, which
allows us to make a comparison of all the observed
slopes to the models in a single color. We have
chosen to correct all the slopes to the F555--F814 vs. F814
slopes of the intermediate redshift sample. This is
done by correcting the observed slope by the change
between the model slopes at the
observed and adopted band-passes. This correction assumes that 
the difference between the model and observed slopes
is independent of wavelength.  It should be noted that this assumption 
is correct only to first order, but should
be essentially true over a small range of wavelengths.
Our main results hinge upon observations of the
highest redshift clusters, where the change in band-passes 
is small (F606--F814 vs. F555--F814), so this correction is
sufficient, and does not effect our conclusions. We have 
corrected to a single color primarily to allow for a 
comprehensive examination of the whole sample.

\paragraph{} The comparison of our observed red sequence slopes to a suite of four models with 
different formation epochs is presented in Figure 3. The overall slope of the red sequence seems
to be consistent with passive evolution for all clusters, implying that the population we have 
sampled have similar evolutionary
histories and time-scales. We note that the observed slopes for the 46 clusters below
redshift 0.35 are all consistent with the models, implying that the mass-metallicity relation used is
reasonable. The observed slopes of the higher redshift clusters are perfectly consistent with a 
high formation redshift, and a conservative limit of $z\geq2$ can be set. Note that up to redshifts 
of z$\approx$0.4,
observations of the slope of the red sequence have little power to distinguish between formation
epochs with z$\geq$1.0.
 The change in slope at z$\leq$0.4
 is exclusively due to blue-shifting of the 
rest-frame band-passes. However, beyond z$\approx$0.5, the expected slopes for the lowest-formation-epoch
models rapidly diverges from the observed slopes. This turnover is a result of the differing 
metallicities in the elliptical population as a function of mass; at
ages younger than $\sim$ 4 Gyr, the
 color evolution of the stellar population in a metal-rich
elliptical is significantly different from that of a
 metal-poorer elliptical (K97; Kodama \& Arimoto 1997). 

\paragraph{}It should be noted that there is another model solution to the observed slopes which
does not require that cluster ellipticals form at $z \geq 2$. This solution requires, however, that
there exist a correlation between formation age and galaxy mass, so that the less massive
elliptical galaxies form later. We have investigated this possible solution in some detail, and
note that while it is possible to produce the observed slopes with this model, it requires
a very specific linear mass vs. formation time relation (amounting to about a 1 Gyr delay in
formation time over the observable six magnitudes of the red sequence), and a small scatter about that relation. 
In the overall parameter space of formation epoch, mass vs.
formation-time relationship slope, and formation-time scatter,
this particular solution occupies a very small volume compared to the high redshift formation scenario.
In the absence of a strong physical motivation for expecting such a specific mass vs. formation-time relation,
we prefer the high-redshift formation solution, with concurrent formation time for galaxies of different
masses.

\paragraph{}
To investigate the sensitivity of our conclusion to cosmological effects, we have constructed
an additional two suites of models with differing cosmologies: 
$H_{0}$=65
, $\Omega_{{\it m}}$=0.1, $\Omega_{\Lambda}$=0.0 and 
$H_{0}$=80, $\Omega_{{\it m}}$=0.2, $\Omega_{\Lambda}$=0.8. The three suites of models
all have similar overall ages, though the precise mapping of the ages to redshift is somewhat
different in each. As can be seen in Figure 4, the differences between the predicted slopes
in different cosmologies are minor, and we may thus parameterize the formation epoch as a
formation redshift. We note that even in a cosmology of a quite different total age, we would still
conclude that at z=0.75 (the highest redshift in our sample), at least
some clusters are older than $\sim$ 4 Gyr. 

\paragraph{}We note that this result does not preclude the formation of some clusters at 
lower redshifts, as our sample may be preselected to examine a particular set of ``old'' objects
(SED98). We may be simply observing the ``oldest'' objects at any particular epoch; it is not
clear that the high redshift clusters in our sample are in fact the antecedents of
the low redshift sample (Kauffmann \& Charlot 1997). SED98 discuss this possibility in some detail , and
note that the results of Kauffmann \& Charlot are derived for a $\Omega$=1 CDM universe, and
that in a low $\Omega$ open universe (Carlberg\etal 1996) clusters are assembled at
significantly higher redshifts. It may also be possible that at z$>0.5$ there is an envelope of
red sequence slopes, where some later forming clusters populate the turnover region. In most reasonable 
cosmologies, we would expect these later forming clusters to be generally less massive. Future observations
of a large unbiased sample of clusters at $0.5<$z$<1.0$ might thus allow us to constrain
the average formation epoch of clusters as a function of mass.

\section{CONCLUSIONS}
\noindent We have demonstrated that the slope of the red sequence, which can be determined 
independent of calibration, can be used to constrain the formation epoch of the early 
type galaxies in the cores
of rich clusters. We have computed the slopes of the red sequences for a total of 50 
clusters, spanning the
redshift range $0<$z$<0.75$.
The data have been reduced and analyzed in a consistent manner for all clusters, in an
effort to reduce systematic variations in the derived photometry. The
low redshift sub-sample uses large ensembles of clusters to strengthen the
red-sequence signal; the sparser high-redshift HST data has been
analyzed morphologically to similarly isolate the red sequence.  We compare these
observations to the latest galactic-wind models for monolithic formation of elliptical galaxies,
and overall find good agreement. The observed slopes show no evidence for 
a turn-over prior to z=0.75, from which we conclude, conservatively,  that at least some clusters
 must have formed at z$>$2.
Finally, we note that the change of the slope of the red sequence with redshift is a potentially
powerful tool to discriminate between formation scenarios for cluster ellipticals, particularly
at redshifts greater than 0.75. A more thorough investigation of the formation history
 of cluster ellipticals
awaits the identification and observation of a significant unbiased cluster sample at these redshifts.

\paragraph{}{\it Acknowledgements.} M.D.G thanks the Natural Sciences and Engineering Research Council of
Canada for support through a PSGA graduate scholarship. O.L.C.
thanks CONACYT-Mexico for
an overseas scholarship for graduate studies at the University of
Toronto and INAOE-Tonantzintla for financial support. T.K. thanks the Japan Society for Promotion of
Science (JSPS) Postdoctoral Fellowships for Research Abroad for supporting his stay
in the Institute of Astronomy, Cambridge, UK. H.Y. acknowledges
support from NSERC of Canada via an operating
grant. We are also grateful to
the referee, J.M. Schombert, and the editor, G. Bothun, for helpful
suggestions which improved the overall quality of the paper and, in
particular, clarified our discussion of S0 contamination.

\vspace{1.0cm}
\centerline{REFERENCES}

\ref {Abraham, R.G., Valdes, F., Yee, H.K.C., \& van den Bergh, S. 1994, ApJ, 432, 75} 
\ref {Arag\'{o}n-Salamanca, A., Ellis, R.S., Couch, W.J., \& Carter, D. 1993, MNRAS, 262, 764}
\ref {Arimoto, N., \& Yoshii, Y. 1987, A\&A, 173, 23}
\ref {Bower, R., Kodama, T., Terlevich, A. 1998, MNRAS, submitted}
\ref {Bower, R., Lucey, J.R., \& Ellis, R.S. 1992a, MNRAS, 254, 589}
\ref {Bower, R., Lucey, J.R., \& Ellis, R.S. 1992b, MNRAS, 254, 601}
\ref {Brown, J.P. 1997, Ph.D. Thesis, University of Toronto}
\ref {Burnstein, D., \& Heiles, C. 1982, AJ, 87, 1165}
\ref {Burnstein, D., \& Heiles, C. 1984, ApJS, 54, 33}
\ref {Castander, F.J., Ellis, R.S., Frenk, C.S., Dressler, A., \& Gunn, J.E. 1994, ApJ, 424, L79}
\ref {Carlberg, R.G. 1984a, ApJ, 286, 403}
\ref {Carlberg, R.G. 1984b, ApJ, 286, 416}
\ref {Carlberg, R.G., Yee, H.K.C., Ellingson, E., Abraham, R.G., Gravel, P.,
 Morris, S.L., \& Pritchet, C.J. 1996, ApJ, 462, 32}
\ref {Carlberg, R.G., Yee, H.K.C., Ellingson, E., Morris, S.L., Abraham, R.G., Gravel, P.,
 Pritchet, C.J., Smecker-Hane, T., Hartwick, F.D.A., Hesser, J.E., Hutchings, \& J.B., Oke, J.B. 1997,
ApJ, 485, L13}
\ref {Dressler, A. 1980, ApJ, 236, 351}
\ref {Dressler, A., Oemler, A., Butcher, H.R., \& Gunn, J.E. 1994, ApJ, 430, 107}
\ref {Dressler, A., Oemler, A., Couch, W.J., Smail, I., Ellis, R.S., Barger, A.,
Butcher, H., Poggianti, B.M., \& Sharples, R.M. 1997, ApJ, 490, 577} 
\ref {Eggen, O.J., Lynden-Bell, D., \& Sandage, A.R. 1962, ApJ, 136, 748}
\ref {Ellingson, E.,\etal 1998, in preparation}
\ref {Ellis, R.S., Smail, I., Dressler, A., Couch, W.J., Oemler, A., Butcher, H., \& Sharples, R.M., 
1997, ApJ, 483, 582 (E97)}
\ref {Holtzman,J.A., Burrows, C.J., Casertano, S., Hester, J.J.,
 Trauger, J.T., Watson, A.M., \& Worthey, G. 1995, PASP, 107, 1065}
\ref {Kauffman, G., \& Charlot, S. 1997, astro-ph/9704148}
\ref {Kodama, T. 1997, Ph.D. Thesis, University of Tokyo (K97)}
\ref {Kodama, T., \& Arimoto, N. 1997, 320, 41}
\ref {Kodama, T., Arimoto, N. Barger, A.J., \& Arag\'{o}n-Salamanca, A. 1997, A\&A, submitted}
\ref {Larson, R.B. 1974, MNRAS, 169, 229}
\ref {Lopez-Cruz, O. 1997, Ph.D. Thesis, University of Toronto (LC97)}
\ref {Lopez-Cruz, O., \& Yee, H.K.C 1998, in preparation}
\ref {Peletier, R.F., Davies, R.L., Davis, L.E., Illingworth, G.D., \& Cawson, W. 1990, AJ, 100, 1091}
\ref {Press, W.H., Teukolsky, S.A., Vetterling, W.T., \& Flannery, B.P. 1992, {\it Numerical
Recipes, The Art of Scientific Computing}, 2nd. edn., Cambridge, Cambridge University Press.}
\ref {Stanford, S.A., Eisenhardt, P.R., \& Dickinson, M. 1995, ApJ, 450, 512}
\ref {Stanford, S.A., Eisenhardt, P.R., \& Dickinson, M. 1998, ApJ,
  492, 461 (SED98)}
\ref {Van Dokkum, P.G., Franx, M., Kelson, D.D., Illingworth, G.D.,
  Fisher, D., Fabricant, D. 1998, astro-ph/9801190} 
\ref {Yee, H.K.C. 1991, PASP, 103, 662}
\ref {Yee, H.K.C., Ellingson, E., Carlberg, R.G. 1996, ApJS, 102, 269}

\vskip 0.75 true in

\centerline{\bf Figure captions}

\noindent{\bf Figure 1a:}The classification plane (concentration index versus magnitude) 
for morphological selection of galaxies in A2390.
All
objects above the cut line (*) have been used in the CMD fitting procedure described in the
text. 

\noindent{\bf Figure 1b:}The color-magnitude diagram for all objects ($\diamond$) and a morphologically-selected subset
(*) of objects in A2390. Note that the restriction of the sample to those objects with a high
central concentration enhances the visibility of the red sequence, particularly at the faint end.

\noindent{\bf Figures 2a-h:}The combined color-magnitude diagrams for the KPNO
sub-sample, consisting of data for 36,385 galaxies (a-e) and the HST morphologically
selected sub-sample, consisting of data for 555 galaxies (f-h). The best-fitting red 
sequences are shown as solid lines.

\noindent{\bf Figure 3:} The comparison of observed red sequence slopes and the predicted slopes
from models with different formation epochs. The cosmology used here, to map the age onto redshift,
is $H_{0}$=50, $\Omega_{{\it m}}$=1.0, and $\Omega_{\Lambda}$=0.0. For
comparison, we also plot the measured slope for the combined CMD of CL0016+1609 \& J1888.16CL
 \& CL0412-65 from E97, and note that our measured slope is perfectly consistent with it.

\noindent{\bf Figure 4:} The predicted slopes for formation models using three different cosmologies
to map age onto redshift.
Note that while the details of the turn-over in the slope for a model at a given formation redshift
changes slightly in different cosmologies, the change is not
significant enough to affect the
conclusion that at least some clusters form at z$>$2.

\begin{deluxetable}{cl}
\tablecaption{The KPNO Sub-Samples}
\tablehead{
\colhead{z-range}&
\colhead{clusters}}
\startdata
0.0232 & Coma(A1656) \nl\hline
0.0414--0.0622 & A85 A154 A168 A407 A671 A957 A1213 A1291 A1795 A1913\nl
& A1983 A2256 A2271 A2399 A2415 A2593 A2626 A2657 \nl\hline
0.0715--0.0852 & A399 A401 A415 A514 A690 A1569 A1650 A1775 A2029\nl
& A2255 A2410 A2420 A2597 A2670 \nl\hline
0.0904--0.1161 & A21 A84 A98 A2244 A2356 A2384 A2440 A2554\nl\hline
0.1303--0.1470 & A629 A646 A1413 A2328 \nl
\enddata
\end{deluxetable}

\begin{deluxetable}{ccccc}
\tablecaption{The HST Sub-Sample}
\tablehead{
\colhead{cluster}&
\colhead{$z$}&
\colhead{F814}  &
\colhead{F606}  &
\colhead{F555}}
\startdata
A2390 & 0.2279$^{a}$& 10500 &\nodata&8400\nl
CL2244-0221 & 0.33$^{b}$& 12600&\nodata&8400\nl
CL0016+1609 & 0.5479$^{a}$& 16800&\nodata&12600\nl
J1888.16CL & 0.560$^{b}$& 16800&\nodata&12600\nl
GHO1322+3114 & 0.755$^{c}$& 20800&10800&\nodata\nl
GHO1322+3027 & 0.757$^{c}$& 32000&16000&\nodata\nl
\tablenotetext{a}{Yee, Ellingson \& Carlberg (1996)}
\tablenotetext{b}{SED98 (1997)}
\tablenotetext{c}{Castander\etal (1994)}
\enddata
\end{deluxetable}

\begin{deluxetable}{ccccc}
\tablewidth{33pc}
\tablecaption{The Combined CMD Sample}
\tablehead{
\colhead{CMD}&
\colhead{$z$}&
\colhead{$\Delta z$}&
\colhead{slope}  &
\colhead{$\Delta$slope}}
\startdata
A1656 & 0.0232  & \nodata &--0.0450&0.0029\nl
KPNO1& 0.0550  & 0.0068 &--0.0539&0.0018\nl
KPNO2& 0.0786 & 0.0042 &--0.0558&0.0021\nl
KPNO3& 0.1002 & 0.0076 &--0.0596&0.0034\nl
KPNO4& 0.1413 & 0.0054 &--0.0662&0.0047\nl
A2390& 0.2279 & \nodata & --0.0370&0.0041\nl
CL2244-0221& 0.330 & \nodata & --0.0423&0.0042\nl
J1888.16CL \& CL0016+1609& 0.551 & 0.008 & --0.0666&0.0067\nl
GHO1322+3114 \& GHO1322+3027& 0.756 & 0.001 & --0.0566&0.0066\nl
\tablecomments{The KPNO sub-sample is divided into 4 groups +
  Coma(A1656) - see Table 1 for details. Slopes are given in the
  original observed bandpass.}
\enddata
\end{deluxetable}

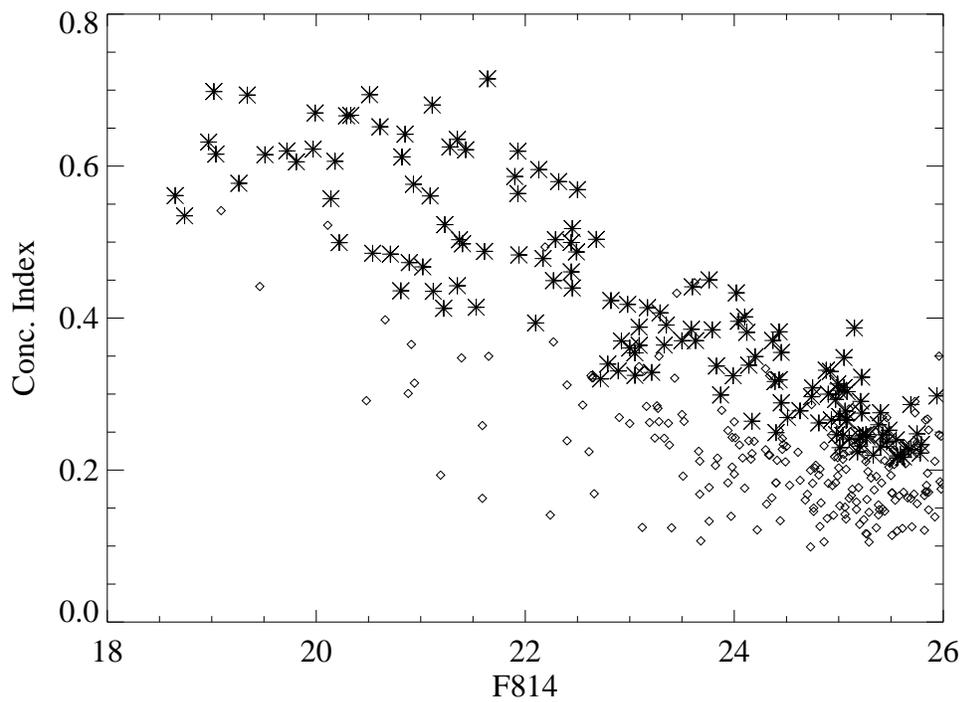
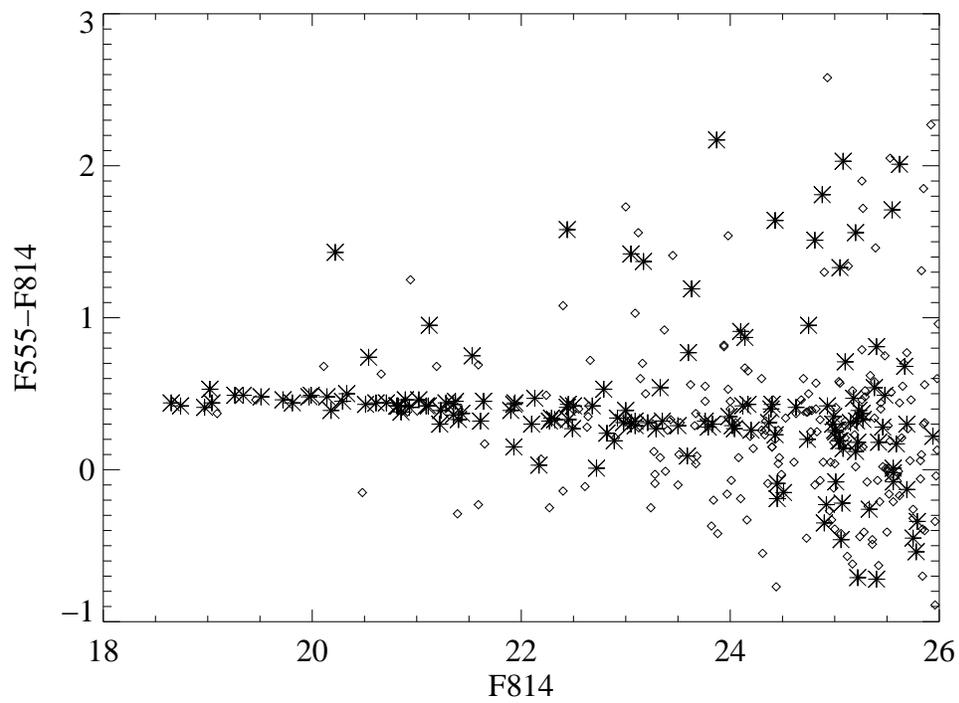
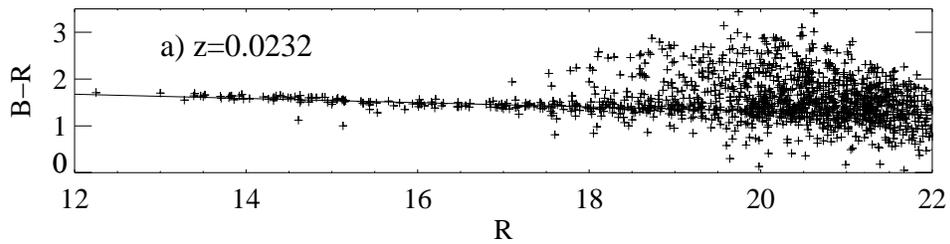
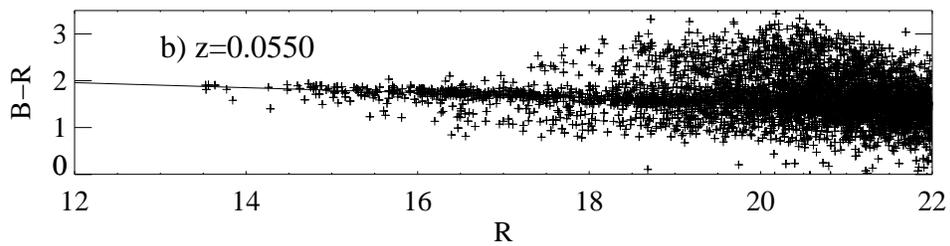
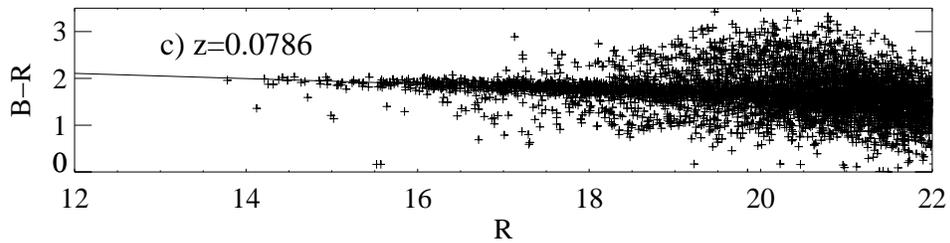
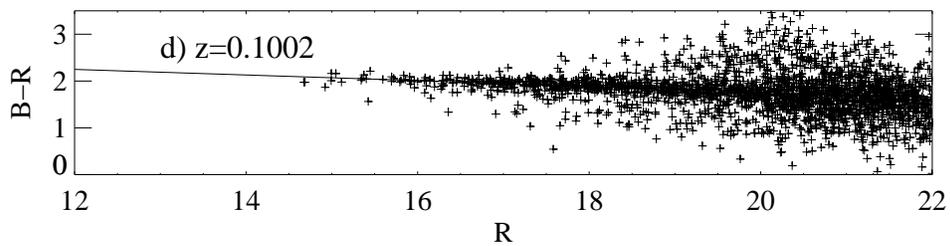
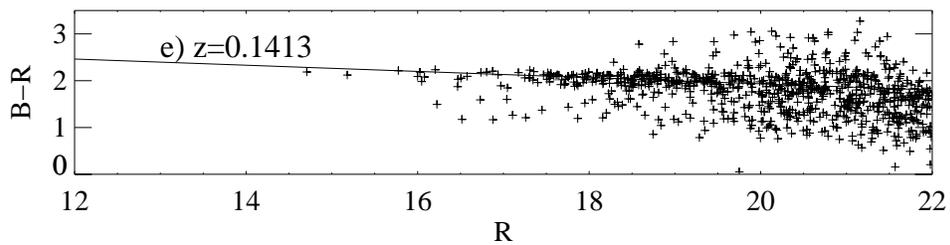
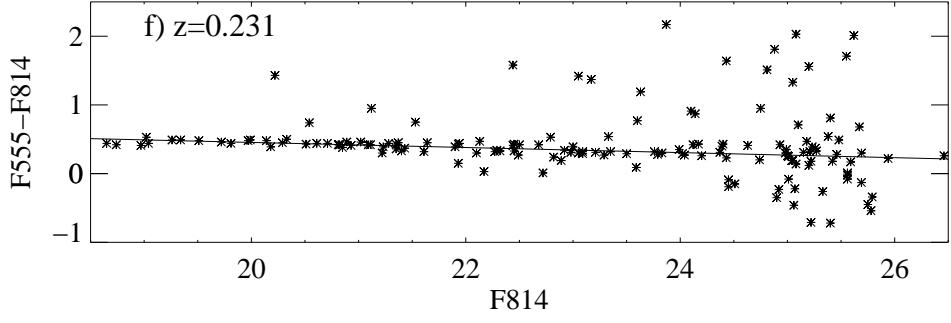
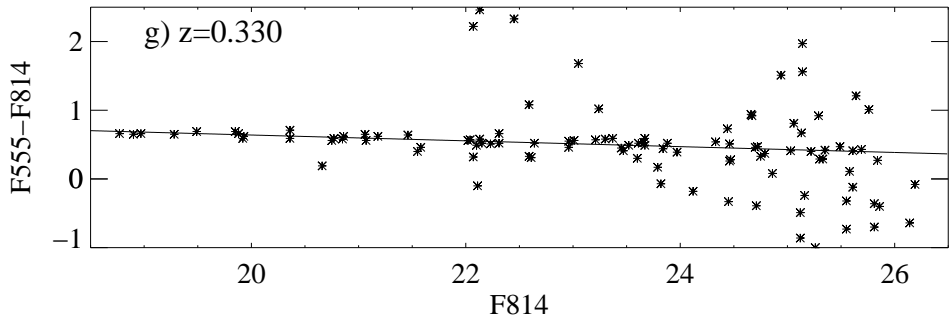
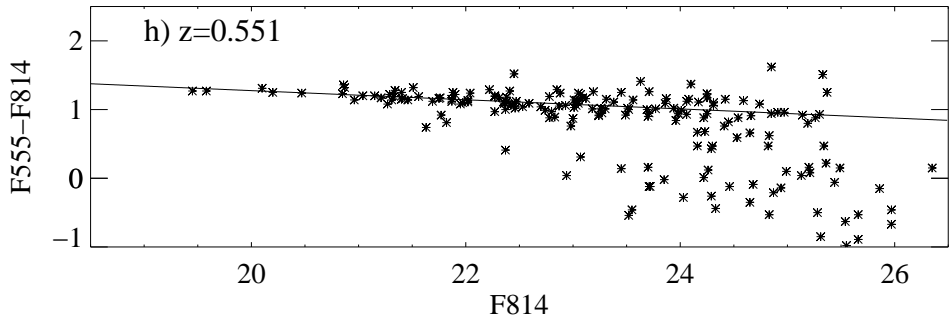
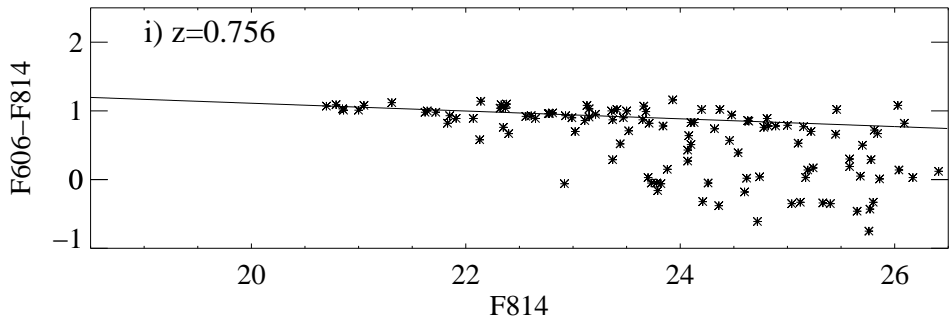
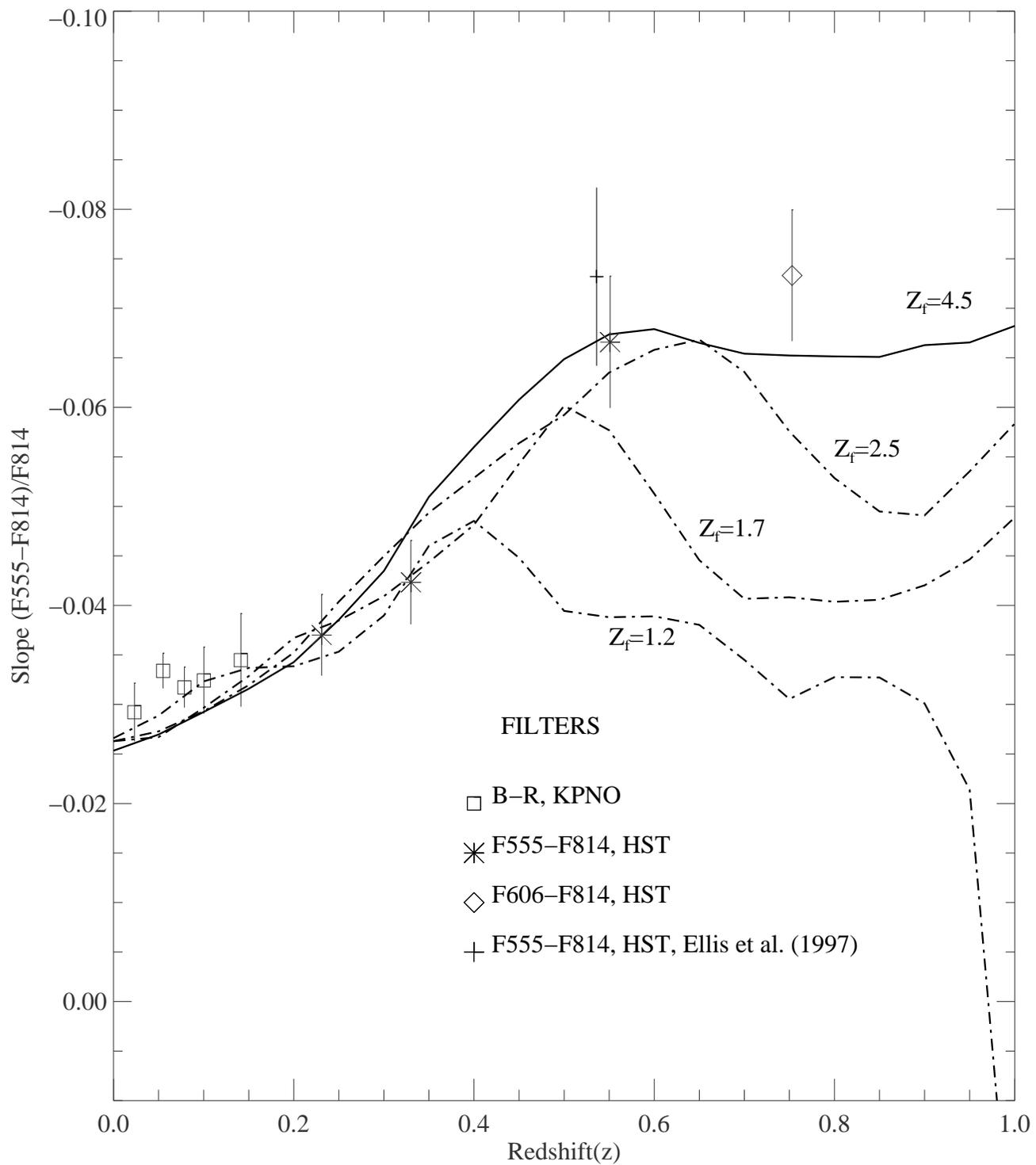
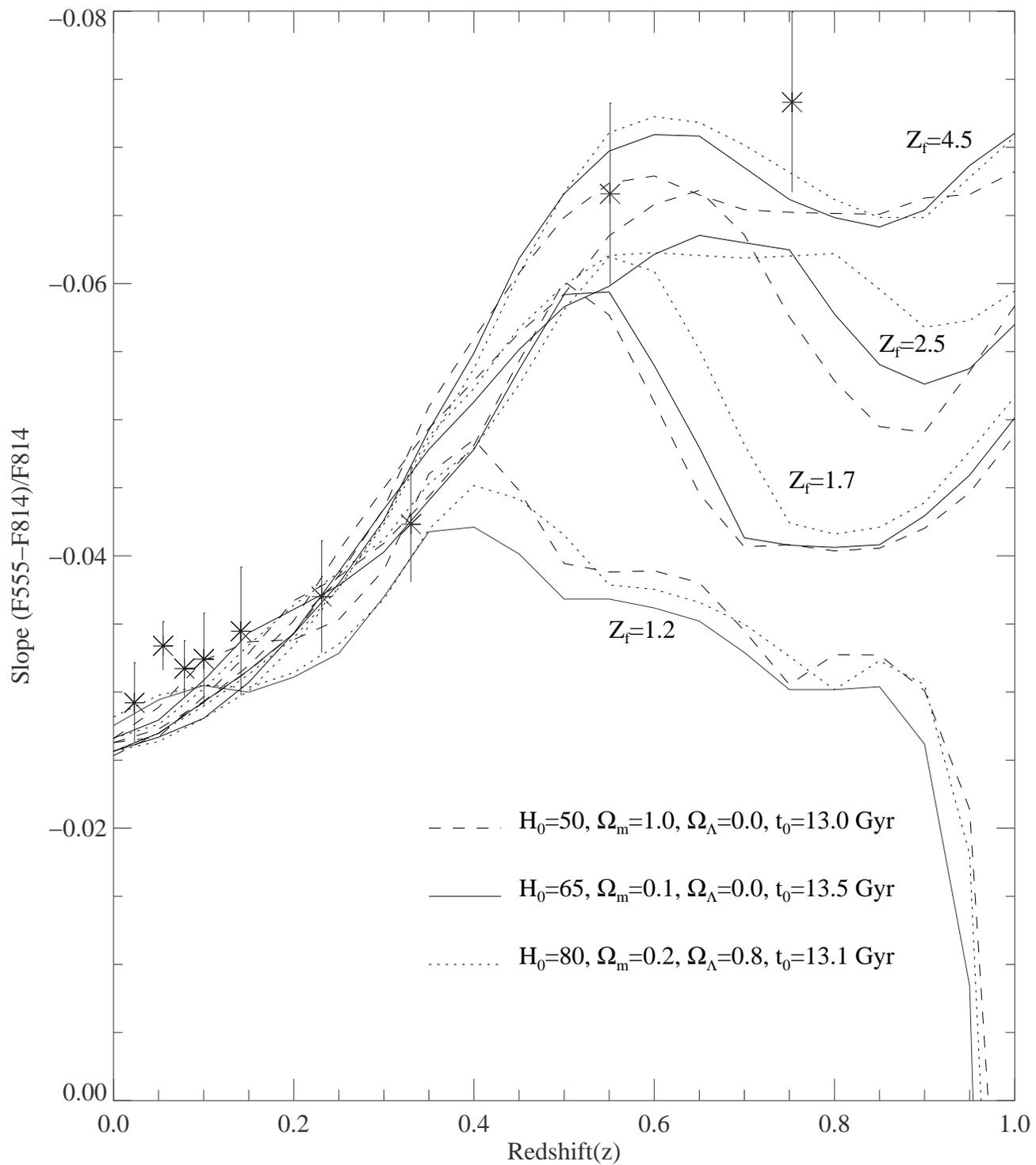
\end{document}